\documentclass[sn-basic]{sn-jnl}

\usepackage{graphicx}%
\usepackage{multirow}%
\usepackage{amsmath,amssymb,amsfonts}%
\usepackage{amsthm}%
\usepackage{mathrsfs}%
\usepackage[title]{appendix}%
\usepackage{xcolor}%
\usepackage{textcomp}%
\usepackage{manyfoot}%
\usepackage{booktabs}%
\usepackage{algorithm}%
\usepackage{algorithmicx}%
\usepackage{algpseudocode}%
\usepackage{listings}%

\usepackage{lmodern}

\begin{document}

\title[R\&D for HCI in Europe]{Advancing European High-Contrast Imaging R\&D Towards the Habitable Worlds Observatory}


\author*[1,2]{\fnm{Iva} \sur{Laginja}}\email{iva.laginja@obspm.fr}

\author[2,5]{\fnm{Óscar} \sur{Carrión-González}}\email{oscar.carrion@obspm.com}

\author[10]{\fnm{Romain} \sur{Laugier}}\email{romain.laugier@kuleuven.be}

\author[5]{\fnm{Elisabeth} \sur{Matthews}}\email{matthews@mpia.de}

\author[6]{\fnm{Lucie} \sur{Leboulleux}}\email{lucie.leboulleux@univ-grenoble-alpes.fr}

\author[9,2]{\fnm{Axel} \sur{Potier}}\email{axel.potier@obspm.fr}

\author[8]{\fnm{Alexis} \sur{Lau}}\email{alexis.lau@lam.fr}

\author[3]{\fnm{Olivier} \sur{Absil}}\email{olivier.absil@uliege.be}

\author[2]{\fnm{Pierre} \sur{Baudoz}}\email{pierre.baudoz@obspm.fr}

\author[4]{\fnm{Beth} \sur{Biller}}\email{bb@roe.ac.uk}

\author[2]{\fnm{Anthony} \sur{Boccaletti}}\email{anthony.boccaletti@obspm.fr}

\author[5]{\fnm{Wolfgang} \sur{Brandner}}\email{brandner@mpia.de}

\author[6]{\fnm{Alexis} \sur{Carlotti}}\email{alexis.carlotti@univ-grenoble-alpes.fr}

\author[5,7]{\fnm{Ga\"el} \sur{Chauvin}}\email{chauvin@mpia.de}

\author[8]{\fnm{Élodie} \sur{Choquet}}\email{elodie.choquet@lam.fr}

\author[14]{\fnm{David} \sur{Doelman}}\email{D.S.Doelman@SRON.nl}

\author[8]{\fnm{Kjetil} \sur{Dohlen}}\email{kjetil.dohlen@lam.fr}

\author[8]{\fnm{Marc} \sur{Ferrari}}\email{marc.ferrari@osupytheas.fr}

\author[16]{\fnm{Sasha} \sur{Hinkley}}\email{S.Hinkley@exeter.ac.uk}

\author[2]{\fnm{Elsa} \sur{Huby}}\email{elsa.huby@obspm.fr}

\author[11]{\fnm{Mikael} \sur{Karlsson}}\email{mikael.karlsson@angstrom.uu.se}

\author[5]{\fnm{Oliver} \sur{Krause}}\email{krause@mpia.de}

\author[9]{\fnm{Jonas} \sur{Kühn}}\email{jonas.kuehn@unibe.ch}

\author[15]{\fnm{Jean-Michel} \sur{Le Duigou}}\email{Jean-Michel.LeDuigou@cnes.fr}

\author[2]{\fnm{Johan} \sur{Mazoyer}}\email{johan.mazoyer@obspm.fr}

\author[12]{\fnm{Dino} \sur{Mesa}}\email{dino.mesa@inaf.it}

\author[14]{\fnm{Michiel} \sur{Min}}\email{M.Min@sron.nl}

\author[6]{\fnm{David} \sur{Mouillet}}\email{david.mouillet@univ-grenoble-alpes.fr}

\author[13]{\fnm{Laurent M.} \sur{Mugnier}}\email{mugnier@onera.fr}

\author[3]{\fnm{Gilles} \sur{Orban de Xivry}}\email{gorban@uliege.be}

\author[1]{\fnm{Frans} \sur{Snik}}\email{snik@strw.leidenuniv.nl}

\author[12]{\fnm{Daniele} \sur{Vassallo}}\email{daniele.vassallo@inaf.it}

\author[8]{\fnm{Arthur} \sur{Vigan}}\email{arthur.vigan@lam.fr}

\author[14]{\fnm{Pieter} \sur{de Visser}}\email{p.j.de.visser@sron.nl}

\affil*[1]{\orgdiv{NOVA}, \orgname{Leiden University}, \orgaddress{\street{Einsteinweg 55}, \city{Leiden}, \postcode{2333 CC}, \country{The Netherlands}}}

\affil[2]{\orgdiv{LIRA, Observatoire de Paris}, \orgname{Université PSL, CNRS, Sorbonne Université, Université Paris Cité}, \orgaddress{\street{5 place Jules Janssen}, \city{Meudon}, \postcode{92195}, \country{France}}}

\affil[3]{\orgdiv{Space sciences, Technologies, and Astrophysics Research (STAR) Institute}, \orgname{Universit\'e de Li\`ege}, \orgaddress{all\'ee du Six Ao\^ut 19c}, \city{Li\`ege}, \postcode{4000}, \country{Belgium}}

\affil[4]{\orgdiv{Institute for Astronomy}, \orgname{University of Edinburgh, Royal Observatory}, \city{Edinburgh}, \postcode{EH9 3HJ}, \country{UK}}

\affil[5]{\orgdiv{MPIA}, \orgname{Max-Planck-Institut für Astronomie}, \orgaddress{Königstuhl 17}, \city{Heidelberg}, \postcode{D-69117}, \country{Germany}}

\affil[6]{IPAG, Univ. Grenoble Alpes, CNRS, 38000 Grenoble, France}

\affil[7]{Laboratoire J.-L. Lagrange, Université Cote d’Azur, CNRS, Observatoire de la Cote d’Azur, 06304 Nice, France}

\affil[8]{Aix Marseille Univ, CNRS, CNES, LAM, Marseille, France}

\affil[9]{University of Bern, Division of Space and Planetary Sciences, Sidlerstrasse 5, 3012 Bern, Switzerland}

\affil[10]{Institute of Astronomy, KU Leuven, Celestijnenlaan 200D, 3001, Leuven, Belgium}

\affil[11]{\orgdiv{Department of Materials Science and Engineering}, \orgname{Uppsala University}, \orgaddress{P.O. Box 35}, \city{Uppsala}, \postcode{751 03}, \country{Sweden}}

\affil[12]{INAF-Osservatorio Astronomico di Padova, Vicolo dell’Osservatorio 5, 35122 Padova, Italy}

\affil[13]{DOTA, ONERA, Université Paris Saclay, F-92322 Châtillon, France}

\affil[14]{SRON - Netherlands Institute for Space Research, Niels Bohrweg 4, 2333 CA Leiden, The Netherlands}

\affil[15]{CNES - Centre National d'Etudes Spatiales, 18 Av. Edouard Belin, 31401 Toulouse, France}

\affil[16]{University of Exeter, Physics Building, Stocker Road, Exeter, EX4 4QL, UK}


\abstract{The Habitable Worlds Observatory (HWO) will enable a transformative leap in the direct imaging and characterization of Earth-like exoplanets. For this, NASA is focusing on early investment in technology development prior to mission definition and actively seeking international partnerships earlier than for previous missions. The ``R\&D for Space-Based HCI in Europe'' workshop, held in March 2024 at Paris Observatory, convened leading experts in high-contrast imaging (HCI) to discuss European expertise and explore potential strategies for European contributions to HWO. This paper synthesizes the discussions and outcomes of the workshop, highlighting Europe's critical contributions to past and current HCI efforts, the synergies between ground- and space-based technologies, and the importance of laboratory testbeds and collaborative funding mechanisms.

Key conclusions include the need for Europe to invest in technology development for areas such as deformable mirrors and advanced detectors, and establish or enhance laboratory facilities for system-level testing. Putting emphasis on the urgency of aligning with the timeline of the HWO, the participants called on an open affirmation by the European Space Agency (ESA) that a European contribution to HWO is clearly anticipated, to signal national agencies and unlock funding opportunities at the national level. Based on the expertise demonstrated through R\&D, Europe is poised to play a pivotal role in advancing global HCI capabilities, contributing to the characterization of temperate exoplanets and fostering innovation across domains.}




\maketitle


\section{Introduction}
\label{sec:introduction}

\subsection{The Vision for Space-Based High-Contrast Imaging (HCI)}
\label{subsec:vision-for-hci}

In recent years, the need for space-based high-contrast imaging (HCI) to directly characterize rocky exoplanets around nearby stars has become a top priority for the international astrophysics community \citep{astro2020,Voyage2050}. Particular interest lies on rocky exoplanets the size of Earth, in such orbits around solar-type stars that might enable surface liquid water -- an area often called the habitable zone -- which requires instruments capable to discern extreme planet-to-star flux ratios (contrasts) of $10^{-10}$. Institutions and agencies across the globe have recognized the importance of characterizing exoplanet atmospheres, particularly those in habitable zones, through deep, high-contrast observations that spatially resolve the exoplanet from its host star.
In the US, the Decadal Survey on Astronomy and Astrophysics 2020 \citep[Astro2020,][]{astro2020} identified the need for a space-based mission enabling the direct imaging of Earth-like exoplanets as its highest priority, leading to the proposal of NASA’s Habitable Worlds Observatory (HWO\footnote{\url{https://habitableworldsobservatory.org/home}}) flagship mission. To ensure the scientific and technical readiness for its future great observatories which includes the ambitious mission of HWO, the decadal survey recommended the establishment of NASA's Great Observatories Mission and Technology Maturation Program (GOMaP\footnote{\url{https://www.greatobservatories.org/newway}}). This nominally decade-long program program aims to develop and mature the essential technologies and retire risk for future NASA mission candidates. As the first entrant to this program, for HWO the early stages of this program focuses on technologies such as starlight-suppression subsystems (coronagraphs), stabilized environments, detectors, deformable mirrors, as well as post-processing algorithms and modeling tools.
In Europe, this ambition is verbalized with the Voyage 2050 roadmap \citep{Voyage2050,snellen2019esavoyage2050white}, which emphasizes mid-infrared observations of exoplanet thermal emissions as critical for high-impact discoveries through a mission project such as the suggested Large Interferometer For Exoplanets (LIFE\footnote{\url{https://life-space-mission.com/}}). These efforts aim to complement HWO, which will focus on exoplanet characterization in reflected light from the ultraviolet (UV) to near-infrared (NIR).
In the meantime, the Nancy Grace Roman Space Telescope, launching in 2027, will carry the Coronagraph Instrument \citep[CGI,][]{kasdinetal2020} that is pivotal for demonstrating the critical technologies needed for HWO, albeit at contrasts about one magnitude shy of the $10^{-10}$. This instrument will not only validate key capabilities like starlight suppression and wavefront control in space, but it could also achieve the groundbreaking milestone of the first-ever direct imaging of an exoplanet in reflected starlight \citep{baileyetal2023}. These developments highlight the growing opportunity for global collaboration and technological innovation in advancing space-based HCI capabilities. With some significant involvement in current HCI-enabling missions as presented in Sec.~\ref{subsec:Europe_current_future}, European researchers are seeking to align their efforts and contribute meaningfully to these initiatives.

In March 2024, a group of European researchers working in HCI for exoplanets organized the “R\&D for Space-Based HCI in Europe” workshop\footnote{\url{https://hcieurope.sciencesconf.org/}} at Paris Observatory, France, to explore how European institutions can contribute to the advancement of space-based HCI. Recognizing Europe’s long-standing contributions to HCI, the primary goal was to exchange, and gather an overview of ongoing activities and already existing cutting-edge projects in European HCI. The workshop brought together experts from various aspects of R\&D for direct imaging techniques, from observations and exoplanet modeling to fundamental technology, simulation and algorithm development, recognizing that advancing this technology requires a broad range of expertise, including space-based instrumentation beyond HCI, and ground-based insights.

This white paper is based on the discussions and conclusions from this workshop. Its objectives were to consolidate European expertise and identify synergies with international projects such as NASA’s HWO. By showcasing Europe’s strengths in this field, the workshop aimed to foster meaningful discussions on Europe’s role in advancing HCI research both towards future space telescopes like HWO, as well as through independent initiatives.

In the remainder of Sec.~\ref{sec:introduction}, we present the main science drivers for space-based direct imaging across a wide wavelength range, and provide an overview of past European direct imaging projects and developments towards HCI from space. In Sec.~\ref{sec:big-missions} we present the goals and current status of the HWO mission and highlight existing technology gaps as well as the missions synergies with LIFE. In Sec.~\ref{sec:European_rd_future_roadmap} we list current involvement of European players in space-based HCI missions, highlight complementarities with ground-based HCI, report on the impact of testing and laboratory facilities, and briefly mention funding structures to enable crucial R\&D activities in Europe on timescales compatible with the milestone requirements for HWO. We give an overall conclusion of the topics discussed during the workshop in Sec.~\ref{sec:conclusions}.

\subsection{Science Drivers for Space-Based HCI}
\label{subsec:science_drivers}

Since the first detections of planetary-mass companions in direct imaging \citep{chauvinetal2004, maroisetal2008, Lagrange2009AProbableGiantPlanet}, the high-contrast imaging technique has probed a population of long-period planets around nearby stars \citep[most recently,][]{nielsenetal2019, viganetal2021}. These are more difficult to detect with the transit and radial-velocity (RV) techniques as they are geometrically biased to shorter periods.
So far, young, self-luminous massive planets have been the main targets for direct imaging due to their relative brightness. However, upcoming instruments will improve the current sensitivity levels by orders of magnitude.
This will unveil a population of colder, lower-mass exoplanets on long-period orbits that remain out of reach to date, and it will provide complementary data on close-in planets accessible by RV as well by providing the first atmospheric characterization for many of these targets.
Already-available instruments such as the coronagraphic modes of the James Webb Space Telescope \citep[JWST,][]{boccalettietal2022, kammereretal2023} provide in-depth characterization of known targets \citep{matthewsetal2024}, but a significant improvement of technical accessibility of shorter orbits will be needed to image faint planets on 1-au orbits around solar-type stars.

Current developments for space-based HCI and nulling interferometry facilities build on work from the early 2000s for mission concepts such as NASA’s Terrestrial Planet Finder - Coronagraph \citep[TPF-C,][]{Traub2006TPF-C} and - Interferometer \citep[TPF-I,][]{Martin2008a}, ESA’s Darwin \citep{Leger2007}, or the Spectro-Polarimetric Imaging and Characterization of Exoplanetary Systems \citep[SPICES,][]{boccalettietal2012}.
These early works already showed that reflected starlight (IR/O/UV) and thermal emission (mid-IR) probe unique, highly complementary physical properties of a planet and its atmosphere \citep[e.g.][]{selsisetal2008}.
This remains relevant with the complementary IR/O/UV and mid-IR approaches of HWO and LIFE \citep{konradetal2022, aleietal2024}.

In reflected starlight, direct-imaging measurements are particularly sensitive to the clouds and aerosols in (exo-)planet atmospheres.
The IR/O/UV range can also probe other phenomena highly relevant to astrobiology, such as the glint of a surface ocean \citep{lustigyaegeretal2018, vaughanetal2023} or the vegetation red edge \citep{saganetal1993, seageretal2005}.
Clouds and aerosols might however hinder an accurate atmospheric characterization as they can mimic or mask other spectral features.
This introduces strong degeneracies in atmospheric retrievals \citep{lupuetal2016, nayaketal2017,damiano-hu2020, carriongonzalezetal2020}. This can be mitigated for example through phase-curve measurements, an approach that has been successful for Solar-System bodies \citep[e.g.,][]{mallamaetal2006, garciamunozetal2014, mayorgaetal2016}, and theoretical work has shown it will also help in characterizing directly imaged exoplanets \citep{nayaketal2017, damianoetal2020, carriongonzalezetal2021b}.
For this, the exoplanet has to be observed as close as possible to both full-phase and transit.
These science observing strategies require observing planets at very small angular separations from their host stars. This makes instrument specifications, such as the small inner working angle and minimum contrast, particularly crucial. As another tool for these goals, optical/near-IR polarimetry can uniquely and unambiguously aid to characterize the atmospheres of exoplanets in reflected starlight and search for spectral features like a surface ocean \citep{trees-stam2019}. Furthermore, it can be used to study planet-forming and debris disks around other stars \citep[e.g.][]{perrinetal2015}, which helps in constraining the properties and abundance of their grains. This is highly relevant to estimate and mitigate the exozodi levels around known exoplanet-hosting stars \citep{erteletal2020}, which is one of the obstacles HWO will face for the direct imaging of low-mass, habitable-zone planets \citep{douglasetal2022}.

Emitted-light observations (mid-IR), at longer wavelengths than the visible, are barely sensitive to aerosols and can therefore measure the temperature profile \citep{konradetal2023} and the radius of a planet \citep{konradetal2022} without being affected by the cloud coverage.
The mid-IR range also has strong spectral bands from molecules that can act as biomarkers or provide context on the habitability of a planet, such as H$_2$O, CO$_2$, CH$_4$, O$_3$ or PH$_3$ \citep{selsisetal2008, aleietal2022, konradetal2022, angerhausenetal2023}.
Combining both spectral ranges would enable a comprehensive characterization of a planet \citep{aleietal2024}. This approach would break the radius-albedo degeneracy, derive the wavelength-dependent albedo, and determine the planet’s energy budget. These insights could then reveal the presence of an eventual greenhouse effect.

In summary, direct imaging will be the technique that enables the atmospheric characterization of cold and temperate exoplanets on long-period orbits, including low-mass habitable-zone planets around Sun-like stars. 
The possibility of combining IR/O/UV and mid-IR data from HWO and LIFE opens the door for a comprehensive characterization of those targets detectable by both instruments, which are expected to be numerous \citep{kammerer-quanz2018, starketal2019, Carrion-Gonzalez2023}.
Preliminary science analysis based on simulated observations can inform the technological development on how to optimize the science return. 
While the exoplanet science discussed above is the prime driver for the mission, HWO will enable many more transformational astrophysics discoveries, acting as a successor of the Hubble Space Telescope (HST) in the IR/O/UV spectral range. 
After the foreseeable degradation of Hubble’s capabilities and potential decommissioning before the 2040s, HWO will keep a unique window open to some of the most prominent astrophysical processes in the Universe, allowing the study of large-scale structures and their evolution through cosmic time.

\subsection{Europe’s Legacy in Space-Based HCI missions}
\label{subsec:european-legacy}

The history of space-based HCI in Europe has been marked by significant milestones and evolving challenges. The coronagraph itself has been invented in Europe \citep{Lyot1939StudySolarCorona}, initially as a means to occult the Sun in order to image its faint corona. After it was adapted to be pointed at solar-system objects and ultimately stars \citep{Vilas1987CoronagraphAstronomicalImaging} in the US, it led to first significant discoveries in the 1980s with the first hints of exoplanetary systems, including the first direct observation of a circumstellar disk around the star $\beta$ Pictoris \citep{Smith+Terrile1984}. Subsequently, HST images revealed debris disks where exoplanets were believed to reside \citep[e.g.,][]{Schneider1999,Schneider2003}.

The early stages of concrete exoplanet discovery, mainly driven by RV and transit methods, further motivated the push for direct imaging technologies. Significant milestones in exoplanet discovery included the first detection of 51 Peg b via radial velocity in 1995 \citep{51Pegb1995}. This was followed by the first observation of an exoplanet in transit \citep[HD 209458b,][]{Charbonneau2000}, shortly followed by the discovery of a planet with the transit method \citep[OGLE-TR-56,][]{Udalski2002,Konacki2003}. These parallel discoveries and early examples show the need for complementary approaches to probe various exoplanet populations. Building on this early stage of exoplanet population statistics studies, we have now entered and are developing a phase of exoplanet characterization that will flourish with future missions.

For direct imaging, it was initially thought that a sufficient signal-to-noise ($S/N$) ratio could be achieved by simply integrating longer due to the quadratic relationship between $S/N$ and integration time $t$ ($S/N \propto \sqrt{t}$) \citep{Nakajima1994, Angel1994}. However, quasi-static speckles, identified in the late 1990s and early 2000s, proved to be a persistent issue \citep{Hinkley2007TemporalEvolution,Cavarroc2006,Boccaletti2003,Bloemhof_2001}, causing a $10^{-4}$ contrast floor in HST. These speckles, caused by slowly evolving wavefront errors, mimick planetary signals, making longer integration insufficient for detecting faint sources. Their evolution is too slow to average out into a speckle halo and too fast to be captured in a PSF reference commonly used to calibrate the speckle spatial distribution, leading to a residual spatial speckle noise floor that is partially removable through careful post processing. Therefore, active suppression of quasi-static speckles at the time of image acquisition is required to effectively treat speckle noise.

The early 2000s saw an important shift in the field of direct imaging with the development of NASA’s TPF mission concepts \citep{Traub2006TPF-C,Martin2008a}. Concepts such as apodized square apertures and hypertelescopes were explored in close collaboration with U.S. teams, organized into dedicated science teams selected by NASA's Jet Propulsion Laboratory (JPL) and including industry partners, with direct involvement of European researchers. Despite the significant technological innovations these studies produced, TPF was ultimately canceled by the U.S. Senate due to funding constraints and technical challenges. In Europe, similar difficulties were faced with the cancellation of the Darwin mission \citep{Leger2007}, which aimed to achieve similar mid-IR nulling interferometry as the TPF-I concept. Although these pioneering concepts did not materialize into missions, they laid the groundwork for future advancements in exoplanet imaging technologies.

Following the cancellation of Darwin and TPF, European efforts continued through initiatives like the Blue Dot Team and the European Planetary Imaging and Characterization Team \citep[EPRAT,][]{Hatzes2010}. Between 2008 and 2010, these teams explored pathways for exoplanet characterization, ultimately recommending the EChO (Exoplanet Characterization Observatory) mission, which later evolved into Ariel, a space-based transit mission focused on atmospheric characterization through spectroscopy \citep{Tinetti2016}. However, both teams also advocated for the importance of direct imaging for the detection and characterization of long-period exoplanets.

In 2010, Europe advanced its independent efforts with the proposal of the SPICES mission \citep{boccalettietal2012,Maire2012Atmospheric}, under ESA’s Cosmic Vision program. Originally submitted as SEE-COAST \citep{Schneider2006seecoast} for the M1 and M2 mission calls, the mission proposed a 1.5-meter off-axis telescope operating in the visible spectrum (450-900~nm), utilizing high-contrast coronagraphs in combination with spectropolarimetric capabilities. Some concepts in SPICES -- for example splitting the light beam per polarization -- are now being considered for future missions like HWO \citep{Feinberg2024}. The mission’s goals included detecting and characterizing exoplanets, distinguishing between super-Earths and mini-Neptunes, and identifying key atmospheric molecules like oxygen and water. However, SPICES was ultimately not selected for the M3 call due to several concerns, largely related to its small aperture and technical immaturity. There was significant overlap with science goals that were expected to be addressed by ground-based facilities such as the Extremely Large Telescope (ELT), using an aperture 25 times larger than SPICES. Additionally, there were doubts about the ability to detect molecular oxygen (O$_2$) with the comparatively low planned spectral resolution of 50--60. Several technical and programmatic challenges further hindered its selection, including concerns about achieving the required pointing stability at the milliarcsecond level, and the complexity of the payload, which involved two deformable mirrors (DMs). Additionally, the polarimetric concept itself posed a significant risk, as it represented a single-point failure due to very little of this technology demonstrated prior to this mission. Ultimately, these factors, combined with challenges in fitting the mission within the M3 schedule and cost cap, led to its rejection.

While the implementation of HCI missions has not been a straight-forward path, they have recently benefited from: (i) the fact that there is indeed a very strong and long-standing science case, (ii) new instrument and mission concepts that significantly helped to consolidate the various approaches by involving a large community, and (iii) the emerging understanding that appropriate preparation is needed to fully face the ambitions of a large HCI mission. On the science side, the progress on exoplanet population statistics and current early characterizations provide important advances to support a future mission with specific science cases focused on nearby systems. On the technology side, previously developed missions, whether deployed or not, are incredibly strong stepping stones for the development of future HCI missions. In particular the large aperture ($\sim$6~m) and investment in technology maturity distinguishes it from earlier efforts. A crucial thing that gives HWO a head start is a clear recommendation in the 2020 Astro Decadal Survey, making it the next flagship mission for NASA. Further, the HWO design will build upon current NASA investments and past missions like JWST (demonstrating the deployment of a large, segmented telescope in space) and Roman (using deformable mirrors in space), as well as long-term tech demonstrations on laboratory testbeds for HCI \citep{Mazoyer2019High}. For all of these, European researchers contributed to the development of key technologies, highlighted in Sec.~\ref{subsec:Europe_current_future}.

This history highlights the cyclical nature of progress in HCI, with the focus shifting between Europe and the U.S. -- nowadays, many of the key direct imaging missions are anchored in U.S.-led efforts. Nevertheless, Europe continues to play an essential role, contributing critical technologies and expertise, and remains poised to shape future direct imaging missions and to participate meaningfully in U.S. initiatives such as HWO.

\section{The HWO mission}
\label{sec:big-missions}

The Habitable Worlds Observatory represents a transformative opportunity in exoplanetary science as it is poised to achieve many of the ambitious science goals outlined in Sec.~\ref{subsec:science_drivers}. With its main science driver being directly imaging Earth-like exoplanets at the $10^{-10}$ constrast level, understanding its objectives and technological requirements is crucial for guiding R\&D efforts in HCI across the globe. The focus on early technology maturation within HWO’s development framework marks a novel and strategic approach that aims to minimize risk and maximize scientific return. However, adapting to this shift in strategy is not straightforward for other agencies and institutions. This approach requires significant adjustments in how mission definition and R\&D is structured, funded, and executed, which can be challenging for organizations accustomed to traditional mission development paradigms.

This section provides an overview of HWO, starting with its mission goals and current development status (Sec.~\ref{subses:mission-goal-status}), followed by a brief discussion of its technology gaps and  development strategies (Sec.~\ref{subsec:technology-development}). The synergies between HWO and the complementary missions concept LIFE are explored in Sec.~\ref{subsec:synergies-life}, highlighting collaborative opportunities in multi-wavelength, direct exoplanet characterization.

\textbf{The discussions during the March 2024 workshop emphasized the need to keep the European community involved in the rapid developments in HWO activities.}

\subsection{Mission goal and status}
\label{subses:mission-goal-status}

The Astro2020 prioritized the study of habitable planets potentially harboring life in the Decadal Survey report \citep{astro2020}, even before the launch of JWST. The report recommends a mission capable of observing at least 25 Earth-like planets orbiting main-sequence stars, aiming to identify one with conditions conducive to life. These target planets should be Earth-like in mass and volume, reside in the habitable zone of Sun-like stars, and possess rocky surfaces capable of supporting liquid water \citep{TheLUVOIRTeam2019}. To achieve these science goals, Astro2020 recommends a space telescope spanning UV to NIR wavelengths ($\sim$100–2500~nm) with a $\sim6$~m inscribed aperture. The telescope's instruments must include coronagraphic imaging and active control to achieve a contrast of $10^{-10}$ at angular separations under 100~mas, coupled with low- to mid-resolution spectroscopy. This mission, now referred to as the Habitable Worlds Observatory \citep[HWO,][]{Feinberg2024,OMeara2024HWO}, has primary objectives to directly detect and characterize Earth-like exoplanets, embedded within broader astrophysical goals including galaxy growth, element evolution, and Solar System studies.

The HWO development plan includes several guiding principles. The first is making the launch date a mission-level requirement, the second is to ensure robotic servicing capabilities at L2 for upgrades and maintenance. It also builds on proven NASA technologies, such as JWST's segmented optics and the Roman Coronagraph. Emphasis is placed on robust scientific and technical margins and maturing critical technologies before entering the development phase, minimizing risk and enhancing mission success.

Large flagship missions often face budget overruns, delays (e.g., JWST), or descope of science objectives. To mitigate these risks, Astro2020 proposed the Great Observatories Mission and Technology Maturation Program (GOMaP\footnote{\url{https://www.greatobservatories.org/newway}}). This decade-long initiative aims to define HWO's concrete science objectives, evaluate instrument capabilities, and explore trade-offs for the mission. GOMaP commenced in 2023, organizing two key groups for HWO: START and TAG. The Science, Technology, and Architecture Review Team (START\footnote{\url{https://www.greatobservatories.org/hwo-start}}) focuses on defining science objectives and determining the observatory and instrument capabilities required. Comprising community scientists from NASA and Europe, START develops models and tools to evaluate expected science outcomes. The Technical Assessment Group (TAG), composed of NASA members, identifies and evaluates mission architecture options, assesses risks, and crafts technology maturation roadmaps. Together, they organize a suite of working groups to ensure readiness for decision-making when Phase A begins. Now dissolved into the official HWO Project Office (PO) established in August 2024\footnote{\url{https://habitableworldsobservatory.org/science?\#h2-c6523d15-73fa-4d21-9a13-56e9c45a9db2}}, START and TAG commenced the work in refining the decadal survey's broad objectives, clarifying top-level requirements like the number of exo-Earths to observe, spectral requirements (wavelength range, resolution, signal-to-noise ratio), and defining ``potentially habitable'' planets, including considerations for their insulation levels, size ranges, and types like super-Earths or water worlds.

By March 2025, the project aims to achieve Concept Maturity Level (CML) 3 \citep{Wessen2013space}. At that stage, a broad exploration of the wide science cases is analyzed against the corresponding impacts in terms of mission concept and critical technological capabilities. The study focus narrows down to identify a few priority science drivers, and the enabling technological challenges which have to be matured in the following years. A refined trade-off study fed by the achieved demonstrations up to TRL~5 at this time will strongly support the identification of a unique mission concept at CML~5 by 2029, optimally addressing the science goals within controlled feasibility and schedule. This important milestone will allow a clear scientific, system and management organization for the actual project development in the following decade for the launch of the mission, equipped with its 1st-generation instrument suite, within the first half of 2040s. 

The ESA Voyage 2050 report underscores the potential for ESA to play a significant role in the next U.S. large flagship mission, specifically HWO. ESA's timely formal commitment to HWO would solidify Europe's involvement in this transformative endeavor.

\textbf{One of the workshop conclusions was that, given the rapid advancements in HWO activities, it would be highly beneficial for ESA to formally express its interest in HWO soon, helping to ensure Europe's timely involvement in this groundbreaking mission.}

\subsection{Technology development}
\label{subsec:technology-development}

As the HWO PO and underlying working groups work to define a suite of science cases for HWO together with corresponding technology requirements, it is imperative to identify the gaps between current technological capabilities and the advancements needed to achieve the outlined scientific objectives. A first technology gap list has been identified at the NASA Astrophysics Project Division (APD\footnote{\url{https://apd440.gsfc.nasa.gov/tech_gap_priorities.html}}) level, following a multi-months study involving various subject matter experts, NASA's Program Offices (among them the Exoplanet Exploration Program, ExEP\footnote{\url{https://exoplanets.nasa.gov/exep/}}) and independent reviewers from the Exoplanet Technology Assessment Committee (ExoTAC). The purpose was to capture the differences between the final instrument specifications and the current state-of-the art.

Prioritization is also made to ensure that the top-ranked technologies receive the greatest attention in terms of funding and personnel at the NASA level, in time to meet the specific missions gate reviews. 
For the gaps determined as the highest interest by the APD (i.e., advancing technologies to close these gaps is judged to be most critical for HWO), we can cite in a non-priority order:
\begin{itemize}
    \item Coronagraphic capabilities in NIR, Vis, and NUV working on complex apertures (segmentation),
    \item coronagraph stability,
    \item sub-Kelvin coolers,
    \item high-reflectivity broadband mirror coatings down to the far UV, and dedicated metrology,
    \item high-throughput integral field spectrographs,
    \item integrated modeling,
    \item mirror technologies for high-angular resolution,
    \item single-photons detection technologies.
\end{itemize}
This list will feed the HWO Technology Maturation Project Office while their technical working groups will refine it depending on new findings and polished science objectives. Inputs on these technology gaps is currently being actively searched for among US-based and international partners.

Currently, NASA is utilizing a series of conceptual observatory designs, referred to as Exploratory Analytic Cases (EACs), to guide the development of the HWO telescope and technology \citep[][Sec.~3 therein]{Feinberg2024}. These EACs serve the purpose of thoroughly exploring the trade spaces for HWO and its instruments, identifying pivotal technology gaps, and assessing architectural options along with critical decision points. By simulating observatory performance and predicting potential scientific outcomes, the EACs offer essential insights into design feasibility. NASA views these EACs as exploratory tools, not finalized concepts, with none intended to become the baseline design for HWO. Instead, they provide a framework to refine strategies and guide subsequent development stages.

For the HCI instrument in particular, coronagraph exploratory cases (CECs) are being studied to assess mass, volume, and power requirements. This effort will incorporate findings from the Coronagraph Design Survey \citep[CDS,][]{Belikov2024}, collecting over 20 coronagraph designs ranging from common masks like Vector Vortex Coronagraphs (VVCs), Apodized Pupil Lyot Coronagraphs (APLCs), Shaped Pupil Coronagraphs (SPCs) to more experimental concepts. The preliminary approach splits the visible and near-infrared light, with a polarization separation for each path, resulting in four channels total \citep{Feinberg2024}. Initial designs include pairs of very large DMs in each channel to improve wavefront control. The next step will focus on evaluating the static raw contrast performance of various coronagraph masks on the EAC. High-level system requirements will be defined through parametric studies of the science yield. An example would be to determine the number of exoEarths that can be characterized as a function of coronagraph performance metrics like raw contrast, post-calibration contrast, and core throughput. These studies will be independent of implementation choices, allowing for a clear comparison between performance requirements and achievable technology. The difference between these will help identify technology gaps and inform the development of a detailed technology maturation roadmap.

One of the conclusions from the workshop discussions was that significant advancements are still required in critical areas such as deformable mirrors, wavefront sensing and control, and detector technologies. Stability, particularly at extreme contrast levels and over long observational durations, remains a persistent challenge that remain to be addressed. \textbf{Given the tight technology development timeline for HWO, R\&D efforts in Europe could make valuable contributions by directly proposing solutions to crucial technology gaps.}

\subsection{Complementarity with the LIFE mission}
\label{subsec:synergies-life}
        
The ESA Voyage 2050 Senior Committee report \citep{Voyage2050} advocates the need for a space mission enabling the characterization of temperate exoplanets in the mid-infrared as a top priority. This comes as a consequence of the radius and atmospheric temperature being key to infer the possibility of surface liquid water. Such a mission would also complement reflected-light observations, as outlined in Sec.~\ref{subsec:science_drivers}. The LIFE initiative\footnote{\url{https://life-space-mission.com/}} seeks to develop the scientific context, the technology, and a roadmap for a mid-infrared space mission that investigates the atmospheric properties of a large sample of terrestrial exoplanets – including 30-50 orbiting within the habitable zone of their host stars. With this, the LIFE mission is set to (a) investigate the diversity of planetary bodies, (b) assess the habitability of terrestrial exoplanets, and (c) search for potential biosignatures in exoplanet atmospheres.

Compared to a visible-light mission like HWO, targeting similar systems at wavelengths roughly 10 times longer prompts the use of apertures 10 times larger than the planned $\sim6$-meter HWO. With interferometry, a 60~m aperture does not have to be a single mirror. Indeed, LIFE would rely on a space-based formation-flying array of collector spacecrafts and an additional beam-combiner spacecraft \citep{Glauser2024}. It would operate over a spectral range of 4-18.5~$\mathrm{\mu m}$ with a spectral resolution of $R\approx 100$. This configuration would allow baselines from tens to hundreds of meters which would enable the required angular resolution. In the regime of around 10\,$\mu m$, which represents the peak thermal emission of an Earth-like planet, the required contrast would be around $10^{-7}$, which is three orders of magnitude more moderate than the reflected light case.

LIFE’s focus on mid-infrared observations complements HWO’s UV to near-infrared capabilities, together offering a combined approach to characterize a wide range of planetary atmospheres, including the detection of biosignatures \citep{Meadows2017a, Schwieterman2018}. This complementary wavelength coverage not only enables comprehensive, multi-wavelength exoplanet characterization but also presents an opportunity to strengthen European technology development efforts aligned with HWO.

LIFE’s advancements in nulling interferometry technologies and innovative instrument architectures highlight significant potential synergies with HWO in critical areas such as starlight suppression, wavefront control, detector development, and data processing techniques. These overlapping technological requirements provide a clear impetus for defining targeted R\&D programs that would address the needs of both missions. Turning these synergies into actionable R\&D initiatives would allow Europe to leverage its expertise in these areas to not only advance LIFE but also significantly contribute to HWO’s success.

\textbf{Workshop discussions emphasized that substantial investment in these technologies—including deformable mirrors, high-precision wavefront sensing and control, and advanced detectors—would benefit both HWO and LIFE.} By formalizing collaborative R\&D initiatives, Europe could directly support HWO’s ambitious goals while simultaneously highlighting LIFE as a potential candidate for ESA’s future mission portfolio.

\section{European R\&D Focus and Future Roadmap}
\label{sec:European_rd_future_roadmap}

The European community has made some crucial contributions to the HCI modes of JWST and Roman. At the same time, Europe has established itself as a leader in ground-based HCI. Expertise in ground-based HCI is becoming increasingly critical for space-based applications, offering complementary technologies and insights that directly support and inform the advancement of future space missions.

This section outlines Europe’s contributions to current and future HCI missions (Sec.~\ref{subsec:Europe_current_future}), explores synergies between space-based and ground-based HCI efforts (Sec.~\ref{sec:synergy_space_ground}), highlights the importance of laboratory and testing facilities (Sec.~\ref{subsec:labs_and_facilities}), and examines institutional participation and funding mechanisms within the European framework (Sec.~\ref{subsec:roadmap_funding_opportunities}). \textbf{The workshop discussions emphasized the importance of leveraging Europe’s unique strengths to contribute to global HCI initiatives, and to further invest in and build upon these strengths, ensuring Europe continues to play a pivotal role in advancing high-contrast imaging technologies.}

\subsection{Europe's Role in Current and Future HCI Missions}
\label{subsec:Europe_current_future}

The European community, together with ESA, has played critical roles in ongoing NASA flagship missions such as JWST and Roman. ESA delivered the optical system of the Mid-Infrared Instrument \citep[MIRI,][]{Wright2015}, which includes four-quadrant phase masks (FQPM) and a Lyot coronagraph \citep{Boccaletti2015,Boccaletti2022}.
MIRI’s coronagraphy facilitated JWST’s first direct imaging discovery of an exoplanet, $\epsilon$ Indi Ab, the first solar-system age exoplanet to be imaged \citep{Matthews2024}.

The Roman Space Telescope, expected to launch by May 2027, will serve as a key testbed for HWO’s high-contrast imaging technologies with CGI \citep{baileyetal2023}. A major European contribution comes from the Laboratoire d’Astrophysique de Marseille (LAM) in France, which provided super-polished off-axis mirrors for the Roman Coronagraph Instrument \citep{Caillat2022SuperPolishedMirrors,Roulet2020OffAxisParabolas}. Based on the stressed-mirror polishing technique (SMP) developed at LAM over several decades, and applied notably to the toroidal mirrors for the main optical path of the SPHERE instrument on the Very Large Telescope (VLT) \citep{Hugot2009AberrationCompensation, Hugot2012ActiveOpticsMethods}, the technique was significantly refined and prototyped on the HiCAT coronagraphic testbed \citep{N'Diaye2014HighcontrastImagerComplex} to enable the manufacturing of the off-axis parabolic shape required for Roman.
In Germany, MPIA designed and manufactured six flight and six engineering models of the Precision Alignment Mechanisms (PAM), which stabilize optical components like mirrors and filters during observations \citep{Krause2025PrecisionAlignment}.
ESA’s contributions include detectors (EMCCDs) for the Coronagraph Instrument, star trackers, batteries, and a new 35-meter antenna in New Norcia, Australia, to support data downlinking.
This involvement led to a Roman participation of the European community in the Core Community Survey (CCS) committees, the Wide Field Instrument (WFI) Project Infrastructure Teams (PIT), and the Coronagraph Community Participation Programme (CCPP).

Roman’s Wide-Field Imager will dominate its observing time, but the Coronagraph Instrument (CGI) will dedicate approximately $\sim$90 days during the first 18 months to demonstrate active wavefront control and speckle-nulling in space, achieving contrasts of $10^{-7}$ or better on-sky. Following this milestone, remaining CGI time will support further scientific and technological demonstrations, which could potentially lead to the first direct imaging detection of an exoplanet in reflected light.
To support Roman science, several working groups have been established, including the Roman CCPP. This group is tasked with providing pipelines for coronagraphic data analysis and planning and executing the technological demonstration observations. ESA has appointed Beth Biller (IfA Edinburgh, UK) and Ga\"el Chauvin (MPIA, Germany) as scientists to represent European interests in the CCPP, ensuring European engagement and reporting back to ESA on group activities.

It is likely that ESA will also play a similarly significant role for the next NASA flagship mission, HWO, although the concrete contribution remains undefined. ESA has appointed three representatives to HWO START, specifically David Mouillet (IPAG, France), Ana Gomez de Castro (UCM, Spain), and Michiel Min (SRON, Netherlands). Their role is to follow developments related to HWO by attending START meetings and related events, report back to the European community and assist the European community in engaging with HWO. Their engagement with the community includes, for example, presenting HWO developments at the R\&D workshop held in March.

\subsection{Synergies with ground-based HCI}
\label{sec:synergy_space_ground}

European efforts in HCI benefit from a wealth of expertise and technology development in both space-based and ground-based observatories. While historically separated due to distinct funding mechanisms and institutional frameworks, recent advances demonstrate that there are significant opportunities to leverage ground-based developments for space missions 
\citep{Pueyo2022CoronagraphicDetection,Potier_2022,Dube2022ExascaleIntegrated,Pogorelyuk2021InformationTheoretical}. This section explores the synergies between space-based and ground-based technologies, as well as other space-based applications such as Earth observation.

For ground-based HCI, the European community has played a central role over the past decades, intimately connected to the pioneer work in the early 90’s, for the development of adaptive optics system, detectors, coronagraphs, observing strategies including differential imaging and data-processing techniques. From a precursor like the COME-ON AO prototype \citep{Kern1989COME-ON} (that would later become the ESO3.6m/ADONIS instrument), new generations of high-contrast imagers like NaCo, SPHERE or ERIS at the VLT \citep{Davies2023ERIS,Lenzen2003,Rousset2003,Beuzit2019}, and soon MICADO, METIS, HARMONI, ANDES and PCS at ELT  \citep{Davies2021MICADO,Brandl2021METIS,Thatte2021HARMONI,Marconi2024ANDES,Kasper2021PCS}, have been developed one after the other to boost performance in terms of contrast, detection and characterization, supported by ongoing R\&D in Europe. The Technology Development Programme of the European Southern Observatory (ESO) plays a critical role in keeping Europe’s ground-based observatories at the forefront of astronomical research. By advancing technologies from low to high TRLs, it ensures their integration into new projects with minimal risk. Unlike conventional procurement processes, technology development at ESO usually requires a collaborative approach between the organization, industry, and research institutes.

The LIFE mission too is being informed by results from a range of ground-based projects, both past and ongoing, like the Keck Nuller \citep{Colavita2009, Serabyn2012}, the LBTI Nuller \citep{Hoffmann2014, Defrere2016, Ertel2018, Ertel2020}, GRAVITY \citep{Lacour2019} and Asgard/NOTT \citep{Defrere2018b, Defrere2022a, Laugier2023}.

One of the main identified challenges in this context is the separation between funding streams for space-based and ground-based technologies. European funding frameworks tend to treat these two domains as distinct, with little coordination between the EU’s Horizon Europe\footnote{\url{https://research-and-innovation.ec.europa.eu/funding/funding-opportunities/funding-programmes-and-open-calls/horizon-europe_en}}, ESA, and national space agencies. This separation often means that R\&D for space and ground-based systems evolves in parallel without direct cross-pollination of technologies. However, ground-based technologies, particularly in adaptive optics and detector development, can have direct applications in space missions, and vice versa. For example, the deformable mirrors developed for the ELT have characteristics—such as high stability and reliability—that are directly applicable to space missions. The same is true for coronagraphs, integral field units and polarimetry.

Despite the challenges, there are opportunities for Europe to take greater ownership of high-TRL technologies for both ground and space missions. DMs are an example of an active area in this context: In the frame of the TANGO project, CNES developed two DMs up to TRL~7 for low-order wavefront control of a 1.5 m class telescope in low-earth orbit \citep{Costes2017ActiveOptics}. One is based on the CILAS monomorph piezoelectric technology with 63 actuators on a 90 mm pupil \citep{Cousty2017MonomorphDMs}, and the second, named MADRAS, uses 24 mechanical actuators on the edge of a 100~mm size mirror \citep{Laslandes2017LastResults}. It was tested by Thales Alenia Space\footnote{\url{https://www.thalesaleniaspace.com/en}} on a full-scale, ground-based telescope demonstrator \citep{Guionie2021FullScale}.
More recently, CNES considered an ALPAO 97-25 DM to very precisely stabilize the line of sight and the WFE of a 40~cm telescope for stratospheric balloons \citep{LeDuigou2023ActiveOptics}.
There is a growing need to adapt these technologies for space-based applications, where factors like stability, reliability, and environmental testing (e.g., vibration resistance in cryogenic environments) become critical.

The European experience in designing, building, testing, calibrating and operating advanced and complex high-contrast imaging instrumentation like SPHERE on the VLT can also be directly applied to new space instrumentation. Lessons learned regarding system-level aspects like broadband performance of optical components (including coronagraphs) \citep{Absil2016ThreeYears} to enable (integral-field) spectroscopy, understanding and mitigation of polarization effects \citep{vanHolstein2020PolarimetricImaging}, and the development of sensing schemes and algorithms, in particular on segmented telescopes \citep{briguglio_2023}, to remove slowly evolving speckles are all relevant to the design of the end-to-end coronagraph instrument for HWO.

\textbf{The workshop discussions emphasized the growing relevance of Europe’s expertise in ground-based HCI for advancing space-based applications. The need to bridge funding and coordination gaps between space and ground-based technologies was a key conclusion, as these synergies offer significant potential to accelerate progress in HCI.}

\subsection{European Laboratory and Testing Facilities}
\label{subsec:labs_and_facilities}

HCI technologies are critical for missions like HWO, necessitating extensive laboratory demonstrations to mature them to the required TRLs. These tests identify potential issues, improve system stability, and validate performance under simulated space conditions. Around the world, several laboratories operate testbeds designed to advance HCI technologies \citep{Mennesson2024CurrentLaboratory}. These testbeds fall into two categories: in-air testbeds and those operated within vacuum chambers. While in-air testbeds are limited to contrasts above $10^{-9}$ \citep{Soummer2024HiCAT,Baudoz2018OptimizationPerformanceMultideformable} due to stability and thermal control limitations, the exact reasons behind these limitations are not yet fully understood. Vacuum testbeds, with their ability to provide better stability and thermal control, are considered essential to reach contrasts on the $10^{-10}$ level. These facilities are thus crucial for maturing and demonstrating HCI technologies required for HWO.

GOMaP aims to advance HCI technologies to TRL~5 by the early 2030s, in preparation for HWO's Phase A. This timeline is relatively compressed, spanning approximately five to six years, and underscores the urgency of development efforts. However, it also presents a strategic opportunity for Europe to contribute to and integrate with these technology maturation initiatives.

Space-based HCI instruments require a broad range of technologies, including coronagraphs, focal-plane wavefront sensing and control, low-order wavefront sensing, machine learning methods, broadband capabilities, spectroscopy, polarimetry, and adaptive optics (AO) stability in space. All these technologies must undergo laboratory demonstrations to validate their feasibility and performance. They are conducted through testbeds, precursor missions, balloons, and complemented by ground-based studies.

Some technologies, such as coronagraphs, are very advanced but still face challenges such as achieving $10^{-10}$ in a laboratory setting with individual designs. Further research is needed to address unresolved challenges, such as integrating coronagraphs with spectroscopy and polarimetry, and optimizing wavefront sensing and control. The only European facility performant enough to work at space-equivalent contrast levels is the \textit{Très Haute Dynamique 2} (THD2) testbed at LIRA in Meudon by Paris, France \citep{Baudoz2018OptimizationPerformanceMultideformable}. Operational for over a decade, it has supported studies for both ground-based and space-based HCI applications \citep{Baudoz2024PolarizatinoEffects,Potier2020_THD2,Galicher2020AFamily,Herscovici-Schiller2018ExperimentalValidationNonlinear,Bonafous2016Development,Mazoyer2014SCC}. Most recently, it installed a mode with appropriate pupil and coronagraphic masks to serve as a testbed demonstrator for the coronagraphic modes of Roman \citep{Laginja2025THD2Roman}. Recent upgrades, including new hardware and control systems, enhance its usage flexibility to test advanced technologies, such as liquid-crystal coronagraphs aiming for $10^{-10}$ contrast. This project \citep[``SUPPPPRESS'',][]{Laginja2024suppppress}, funded by ESA's Directorate of Technology, Engineering and Quality (TEC), initially aims to progress results to the $10^{-9}$ level on THD2, eventually aiming for $10^{-10}$ when transferred to a vacuum testbed, currently unavailable in Europe.
The contrast difference between $10^{-9}$ and $10^{-10}$ is directly influenced by the operational environment. An in-air testbed like THD2 can be used to develop and test space-based HCI components up to a $10^{-9}$ contrast level. This preliminary step is complex and requires suitable infrastructure. However, it optimally prepares components for the subsequent $10^{-10}$ investigation stage in a vacuum environment. This step-wise approach allows for efficient resource allocation, as $10^{-10}$ testbeds are freed from initial development work that can be conducted in non-vacuum settings.

NASA’s High-Contrast Imaging Testbed (HCIT) laboratory at JPL has been a cornerstone for addressing these issues. HCIT hosts two vacuum testbeds dubbed the Decadal Survey Testbeds (DST) 1 and 2 \citep{Noyes2023TheDSTtwo,Seo2019TestbedDemonstration,Patterson2019DesignDescription} and several smaller testbeds, all working on key technologies for HCI. In 2007, JPL demonstrated $10^{-10}$ contrast using a Lyot coronagraph in a 2\% bandwidth \citep{Trauger2007}. This experiment represented a major milestone for HCI, but it also revealed the complexity of achieving the necessary stability for long-duration observations. By 2023, fifteen years later, the same contrast was achieved in 20\% broadband, showcasing substantial progress \citep{Allan2023DemonstrationOfCoronagraph}. Other notable testbeds advancing space-based HCI technologies, not managed by NASA but all located in the US, include HiCAT (High-contrast imager for complex aperture telescopes) at STScI in Baltimore, U.S. \citep{Soummer2024HiCAT}, and SCoOB (Space Coronagraph Optical Bench) at University of Arizona in Tucson, U.S. \citep{vanGorkom2022TheSpaceCoro}.

As HWO development advances, U.S. efforts increasingly focus on building new vacuum testbeds to conduct system-level evaluations. However, in-air testbeds remain valuable for addressing open questions about HCI systems. They are easier to build, maintain, and operate, allowing significant progress at contrasts of $10^{-8}$ to $10^{-9}$ before transitioning to vacuum environments to reach $10^{-10}$ contrast. This flexibility enables rapid prototyping and refinement of key technologies.

For example, results from JPL and HiCAT emphasize the distinction between raw contrast, achieved under ideal observing conditions, and post-processed contrast, which can ease laboratory demonstration requirements but demands higher system stability \citep{Redmond2024ExoplanetDetection}. Instabilities during observations can significantly reduce the effectiveness of post-processing. Questions like this can easily be addressed on in-air testbeds to proceed rapidly on critical technologies without the immediate need for complex vacuum testbeds. 

Workshop participants interpreted that one single NASA facility seems sufficient to test and validate one or a few selected HCI systems in their final stages. However, it will not allow the exploration of a range of R\&D variations, such as comprehensive investigations of system studies of differing control algorithms on distinct coronagraph concepts and linked to other instrument parts like spectrographs.
With so much work required and a tight 5-6 year timeline for HWO’s Phase A preparation, relying on JPL’s resources alone seems to be likely to create bottlenecks in advancing critical HCI technologies for HWO.
Additionally, the Roman testbed, once a vital part of JPL’s HCIT laboratory, has since been decommissioned, removing a hands-on facility to support the only mission that will fly before HWO. The points discussed above underscore the opportunity for establishing a European testbed. Even with an extended development timeline, participants noted that such a testbed could align with HWO’s in-situ servicing schedule, ensuring European researchers’ ability to contribute effectively to the mission’s goals.

Workshop discussions highlighted both the feasibility and challenges of establishing a dedicated European testbed, with a particular focus on the potential for a vacuum facility. While this remains the ideal solution, participants noted that a high-performance in-air testbed should also be considered as a big gain. Developing a new testbed would require significant time, funding, and technical expertise. However, lessons from facilities like THD2 and HiCAT, as well as the SCoOB testbed in Arizona — which achieved $10^{-9}$ contrast in vacuum within five years of start of the project \citep{Ashcraft2022TheSpaceCoro} — demonstrate the potential for leveraging existing knowledge to accelerate progress. The success of these testbeds relied heavily on the involvement of mechanical, vacuum, and control engineers.

Funding a testbed, however, in vacuum or in air, without a direct link to a specific mission or instrument remains a major hurdle, as funding agencies prioritize initiatives tied to concrete projects. Workshop participants emphasized the need for a clear strategy if Europe’s space-based HCI expertise is to be maintained and expanded. Formalizing this strategy in a future white paper could help define priorities, increase TRLs for key technologies, and align European efforts with global developments.

Installing a new vacuum testbed in Europe should be thoroughly examined, as it represents the most effective way to simulate space conditions for HCI technologies. However, \textbf{investing in a high-quality in-air testbed provides a low-risk, high-gain alternative, especially for supporting lower-TRL technology development until a vacuum environment is truly needed}. In-air setups are faster to implement, require less infrastructure, and are highly effective for advancing coronagraphy and wavefront control. THD2, the only European testbed providing performance levels comparable to U.S. facilities, exemplifies this approach.

\subsection{Institutional Participation, Funding, and Collaboration Opportunities}
\label{subsec:roadmap_funding_opportunities}

The international collaboration on missions such as Roman and JWST has been led by well-established institutions, including the European partners ESA, CNES (France), and the Max Planck Institute for Astronomy (MPIA) in Germany. These institutions have a long-standing  involvement in space science. However, for other countries with emerging or less robust space programs, direct involvement in large international missions can be challenging.

In countries like Belgium, for example, there is no formal space agency. Instead, Belgium relies on the Belgian Science Policy Office (BELSPO), which traditionally only funds projects after they have already been adopted, limiting prospective activities in R\&D.

In contrast, Spain has recently established a national space agency, but it remains closely tied to the Ministries of Science and Defense. Spain has a long history of collaborating with ESA, and national funding is available through government calls, although the agency's scope is still evolving.

The Netherlands Space Office is the Dutch governmental agency that develops and realizes the national space policy, in close connection to ESA. It actively funds low-TRL R\&D efforts and consequent instrument development by consortia of academic institutes (SRON, the Netherlands Institute for Space Research, and university institutes like Leiden Observatory and NOVA), in collaboration with Dutch industry (e.g., Airbus DSNL, TNO, cosine, Lionix).

In the UK, there is a dedicated space agency, but funding for astronomy is primarily handled by the Science and Technology Facilities Council (STFC), which focuses more on ground-based activities. However, UK involvement in space-based astronomy can be supported by ESA, and there are some major projects underway, such as the new Space Park in Leicester, which may serve as a potential hub for future HCI-related developments.

In Sweden, the Swedish National Space Agency (SNSA) supports research projects with a technical focus where the results are to be used in space applications. The budget is limited (about 600,000 EUR) but it could for instance cover the development and fabrication of new types of vortex coronagraphs for space-based telescopes. However, this call only opens every third year (the last one was 2023), although there are ongoing discussions to open a new call in 2025 already.

In Switzerland, the Swiss Space Office (SSO) can provide funding for engineering and equipment through PRODEX grants, but there is a stipulation that at least 50\% of the funds must go to local industry. This creates both opportunities and constraints for Swiss researchers working on space-related projects, such as the NICE experiment funded for the LIFE mission.

Workshop participants concluded that in many of these cases, national space agencies and institutions look to ESA for guidance. With GOMaP, NASA started soliciting international inputs to HWO technology developments, including closing existing technology gaps in HCI, early in the mission timeline. There has been a wide international response, however, such activities often lack the appropriate resources to support them.
\textbf{
Without a clear ESA roadmap that includes HWO, it can be difficult for smaller countries to justify investments in HCI technologies. Importantly, ESA involvement serves as a crucial signal to national agencies for prioritizing such investments.} This ``chicken and egg" problem -- where national agencies wait for ESA, and ESA in turn reacts to national interest -- makes it challenging to secure early-stage funding for technologies that are not yet tied to concrete missions. \textbf{Workshop participants noted that HWO’s new project paradigm places strong emphasis on both early technology development as well as encourages early national and international engagement, highlighting the need for Europe to align its strategies and investments with this new approach.}

Considering these constraints, the workshop participants encourage European researchers to push for space-based R\&D projects both at the national and international levels. Many agencies will support technological advancements if they align with global efforts, even if they are not yet part of an officially endorsed mission. For example, the European Research Council (ERC) is relatively insulated from programmatic politics and remains a key funding source for exploratory research. ERC proposals are assessed on their scientific merit, and as long as the outcomes align with advancing technical understanding, ERC funding can be secured without needing to tie the project to a specific mission.

\section{Summary and Conclusions}
\label{sec:conclusions}

HWO is advancing rapidly, and the participants of the ``R\&D for Space-Based HCI in Europe'' workshop held in Paris in March 2024 highlighted the urgency for Europe to align its R\&D efforts to remain a significant contributor to global HCI initiatives. The primary goals of the workshop were to consolidate European expertise and foster collaboration on HCI technologies. Immediate outcomes included plans for a follow-up workshop in 2025 in Germany\footnote{\url{https://hcieurope-mpia.sciencesconf.org/}} and the creation of a Slack workspace to enhance communication within the European community working on R\&D for HCI.

The tight timeline for HWO’s technology readiness necessitates immediate action and targeted investments if a European involvement is anticipated. Participants believe that it would be a great programmatic advantage for ESA to formally endorse HWO, even without committing any funding, to signal national agencies to prioritize related projects. Investing in high-TRL technologies such as deformable mirrors, wavefront control, coronagraphs, polarimetric components, and detectors was seen as vital to addressing gaps for HWO and advancing Europe’s capabilities even beyond this particular mission. The limited number of European testbed facilities was highlighted as a constraint, and establishing or upgrading these facilities, including the potential for a vacuum testbed or a high-quality in-air testbed, was strongly recommended.

In conclusion, the workshop underscored the importance of a coordinated European roadmap that integrates national and international efforts to ensure alignment with global HCI developments. By leveraging its unique strengths, Europe can continue to play a critical role in advancing exoplanet characterization and maintaining leadership in HCI technologies.

\backmatter

\bmhead{Acknowledgements}

This work was supported by the Action Spécifique Haute Résolution Angulaire (ASHRA) of CNRS/INSU co-funded by CNES.
I.L. is supported by the European Space Agency (ESA) under the tender number TDE-TEC-MMO AO/1-11613/23/NL/AR in the context of the ``SUPPPPRESS'' project. I.L. acknowledges partial support from a postdoctoral fellowship issued by the Centre National d’Etudes Spatiales (CNES) in France.
This work has received funding from the Research Foundation -- Flanders (FWO) under the grant number 1234224N.
E.C. and A.L. acknowledge funding from the European Union under the European Union’s Horizon Europe research and innovation programme (ERC Grant No.~101044152). O.A. and G.O.X. acknowledge funding from the European Union under the European Union’s Horizon 2020 research and innovation programme (ERC Grant No.~819155). Views and opinions expressed are however those of the author(s) only and do not necessarily reflect those of the European Union or the European Research Council Executive Agency. Neither the European Union nor the granting authority can be held responsible for them.
O.A. is a Senior Research Associate of the Fonds de la Recherche Scientifique – FNRS.


\bibliography{bibliography}

\end{document}